\documentstyle[12pt]{article}
\advance \topmargin by -\headheight
\advance \topmargin by -\headsep

\evensidemargin \oddsidemargin
\begin{document}

\topmargin -.6in

\def\rf#1{(\ref{eq:#1})}
\def\lab#1{\label{eq:#1}}
\def\br{\begin{eqnarray}}
\def\er{\end{eqnarray}}
\def\be{\begin{equation}}
\def\ee{\end{equation}}
\def\lb{\lbrack}
\def\rb{\rbrack}
\def\({\left(}
\def\){\right)}
\def\v{\vert}
\def\bv{\bigm\vert}
\def\lskip{\vskip\baselineskip\vskip-\parskip\noindent}
\relax
\newcommand{\nit}{\noindent}
\newcommand{\ct}[1]{\cite{#1}}
\newcommand{\bi}[1]{\bibitem{#1}}
%
\def\a{\alpha}
\def\b{\beta}
\def\c{\chi   }
\def\d{\delta}
\def\D{\Delta}
\def\eps{\epsilon}
\def\g{\gamma}
\def\G{\Gamma}
\def\grad{\nabla}
\def\h{ {1\over 2}  }
\def\hc{\hat{c}}
\def\hd{\hat{d}}
\def\hg{\hat{g}}
\def\hp{ {+{1\over 2}}  }
\def\hm{ {-{1\over 2}}  }
\def\k{\kappa}
\def\l{\lambda}
\def\L{\Lambda}
\def\m{\mu}
\def\n{\nu}
\def\o{\over}
\def\O{\Omega}
\def\p{\phi}
\def\pa{\partial}
\def\pr{\prime}
\def\ra{\rightarrow}
\def\rh{\rho}
\def\s{\sigma}
\def\t{\tau}
\def\th{\theta}
\def\ti{\tilde}
\def\wti{\widetilde}
\def\inte{\int dx }
\def\xb{\bar{x}}
\def\yb{\bar{y}}
\def\tr{\mathop{\rm tr}}
\def\Tr{\mathop{\rm Tr}}
\def\partder#1#2{{\partial #1\over\partial #2}}
\def\ds{{\cal D}_s}
\def\wtwo{{\wti W}_2}
\def\lie{{\cal G}}
\def\alie{{\widehat \lie}}
\def\dlie{{\cal G}^{\ast}}
\def\elie{{\widetilde \lie}}
\def\edlie{{\elie}^{\ast}}
\def\hlie{{\cal H}}
\def\wlie{{\widetilde \lie}}
\def\rlx{\relax\leavevmode}
\def\inbar{\vrule height1.5ex width.4pt depth0pt}
\def\IZ{\rlx\hbox{\sf Z\kern-.4em Z}}
\def\IR{\rlx\hbox{\rm I\kern-.18em R}}
\def\IC{\rlx\hbox{\,$\inbar\kern-.3em{\rm C}$}}
\def\one{\hbox{{1}\kern-.25em\hbox{l}}}

\begin{titlepage}

\vskip .6in

\begin{center}
{\large {\bf Comments on Two-Loop Kac-Moody Algebras}}
\end{center}

\normalsize
\vskip .4in

\begin{center}

{L.A. Ferreira\footnotemark
\footnotetext{Supported in part by CNPq; e-mail: 47553::LAF, 47553::JFG}},
J.F. Gomes$^{\,1}$ and A.H. Zimerman$^{\,1}$

\par \vskip .1in \noindent
Instituto de F\'{\i}sica Te\'{o}rica-UNESP\\
Rua Pamplona 145\\
01405 S\~{a}o Paulo, Brazil
\par \vskip .3in

\end{center}

\begin{center}
{ A. Schwimmer\footnotemark
\footnotetext{On leave from Department of Physics, Weizmann Institute, Rehovot,
76100 Israel}$^{,}$\footnotemark \footnotetext{Supported by FAPESP-Brazil and
BSF}}

\par \vskip .1in \noindent
SISSA and INFN, Sezione de Trieste\\
2-4 via Beirut\\
Trieste 34014, Italy\\
\par \vskip .3in

\end{center}

\begin{center}
{\large {\bf ABSTRACT}}\\
\end{center}
\par \vskip .3in \noindent

It is shown that the two-loop Kac-Moody algebra is equivalent to a two
variable loop algebra and a decoupled $\beta$-$\gamma$ system. Similarly WZNW
and CSW models having as algebraic structure the Kac-Moody algebra are
equivalent to an infinity of versions of the corresponding ordinary models and
decoupled abelian fields.

\end{titlepage}

Infinite dimensional extensions of the chiral algebras appearing in Conformal
Field Theories received a lot of attention recently. The general feature of
these extensions is the dependence of the currents on an additional variable
besides $z$, the chiral coordinate on the 2-D world sheet. These new algebras
\cite{flor} include the different versions of $W_{\infty}$\cite{bakas,pope} and
the two loop Kac-Moody (KM)\cite{toda,mai,morozov,bmz,W}.

One version of the two loop K-M algebra was introduced in ref.\cite{toda} in
the context of a WZNW type of model having as target space an infinite
dimensional group manifold  which, under a Hamiltonian reduction procedure,
gives rise to the conformal affine Toda model (CAT)\cite{babo}.   Under a
similar line of reasoning a Chern-Simons theory associated to the same
algebraic structure was also considered in ref.\cite{bmz}.

For the $W_{\infty}$ type algebra  it was pointed out by Witten\cite{witten}
that there is a natural geometric interpretation in terms of symplectic
diffeomorphisms of the cotangent bundle $T^*\Sigma$ of a Riemann surface
$\Sigma$. The second variable is therefore intrinsically linked to $z$. Such an
interpretation is lacking for two-loop KM algebras. We suspect that in this
case the two variables can be decoupled.  The main result of this paper is to
show that this is indeed true. The two-loop KM algebra can be decomposed by a
redefinition of currents into a two variable loop algebra, with two central
extensions, and a $\beta$-$\gamma$ system.

As a reflection of this algebraic fact the WZNW and CSW models based on KM
algebras are shown to be equivalent to an infinite number of copies, labeled by
an additional variable, of the ordinary (based on Lie algebra) versions of the
corresponding models and decoupled sets of abelian fields.

Let us start with the algebraic treatment.  The two-loop KM algebra is defined
by the following commutation relations\cite{toda}
\br
\lb J^m_a (x) \, , \, J^n_b (y) \rb &=& i f_{ab}^c J_c^{m+n}(x) \delta
(x - y)  + i {k\o {2 \pi}} g_{ab} \pa_x  \d (x - y)  \delta_{m,-n}
+ C (x) \d (x - y)  g_{ab} m \delta_{m,-n} \phantom{......}
\label{eq:kma} \\
\lb D (x) \, , \, J^m_a (y) \rb &=& m J^m_a (y) \d (x -y)
\label{eq:kmb} \\
\lb C (x) \, , \, D (y) \rb &=& i {k\o {2 \pi}} \pa_x  \d (x - y)
\label{eq:kmc} \\
\lb C (x) \, , \, J^m_a (y) \rb &=& 0 \label{eq:kmd}
\er
where $f^c_{ab}$ are the structure constants of a finite semisimple Lie
algebra $\lie$, while $g_{ab}$ is the Killing form of $\lie$
($g_{ab} = \Tr (T_a T_b)$, $T_a$ being the generator of $\lie$) and
$n,m \in \IZ$.

We will consider (\ref{eq:kma})-(\ref{eq:kmd}) as the algebra of equal time
commutators of the currents with the spatial variable $x$ being on a circle of
radius 1, and the currents satisfying the periodic boundary conditions
$J^m_a(x + 2 \pi) = J^m_a(x)$, $C(x + 2 \pi) = C(x)$ and $D(x + 2 \pi) = D(x)$.

\nit Using the fact that $C(x)$ commutes with $J^m_a(x)$ we can define a new
set of currents by
\be
\tilde{J}^m_a(x) = J^m_a(x) exp\( {m\o {k}} \sum_{r \neq 0} {C_r \o
r}e^{- i r x}\) \label{newcur}
\ee
where $C_r$ are the modes of $C(x)$, i.e.
\be
C(x) = {1 \o {2 \pi}} \sum_{r=-\infty}^{\infty} C_r e^{-irx}
\label{expC}
\ee
The zero mode $C_0$ is not included in the transformation (\ref{newcur}) in
orde
   r
to respect the periodicity properties of $\tilde{J}^m_a(x)$. Notice that the
transformation (\ref{newcur}) is defined on a given representation and not on
the abstract algebra. Using
(\ref{eq:kma})-(\ref{eq:kmd}) and the identity
\be
f(x) \pa_x \d (x-y)  = f(y) \pa_x \d (x-y) - \pa_y f(y) \d (x-y)
\label{id}
\ee
one obtains
\br
\lb \tilde{J}^m_a (x) \, , \, \tilde{J}^n_b (y) \rb &=& i f_{ab}^c
\tilde{J}_c^{m+n} (x) \delta (x - y)  + i {k\o {2 \pi}} g_{ab} \pa_x  \d (x -
y)  \delta_{m,-n} + {C_0 \o {2\pi}} \d (x - y)  g_{ab} m \delta_{m,-n}
\phantom{......}
\label{kma} \\
\lb D_0 \, , \, \tilde{J}^m_a (x) \rb &=& m \tilde{J}^m_a (x)
\label{kmb} \\
\lb D_r \, , \, \tilde{J}^m_a (x) \rb &=& 0 \quad\qquad\;\; r\neq 0
\label{kmc}
\er
where $D_r$ are the modes of $D(x)$, i.e. , $D(x) = {1 \o {2 \pi}}
\sum_{r=-\infty}^{\infty} D_r e^{-irx}$. $D_0$ is the operator measuring the
``momenta'' $m$ in an usual KM algebra\cite{algebra}. Formulae
(\ref{kma})-(\ref{kmc}) show that the $C$-$D$ system is decoupled from the
$\tilde{J}'s$.Introducing $\bar x$ as a conjugate variable to $m$,$\tilde{J}
(x,\bar x)$ obeys an algebra (Eq.(8)) completely symmetric in $x,\bar x$
. $C_0$ commutes with all the operators, i.e., acts like a second
central extension in the algebra of $\tilde{J}'s$.

For certain classes of representations one of the central extensions can be
removed as we now show.  In order to do that we have to make specific
assumptions about the groups and classes of representations. We study algebras
based on a {\em compact} Lie group and we limit ourselves to {\em unitary,
highest weight representations} for the $\tilde{J}^m_a (x)$ currents (we will
discuss the possibility for the decoupled $D-C$ system afterwards). Decomposing
$\tilde{J}^m_a (x)$  in its modes as $\tilde{J}^m_a (x) = {1 \o {2 \pi}}
\sum_{r=-\infty}^{\infty}
\tilde{J}^{m,r}_a  e^{-irx}$, we see from (\ref{kma}) that we can require the
highest weight state to be annihilated by all the operators corresponding to
$(m,r)$ in a half plane limited by $m C_0 + r k > 0$. Then we can use a slight
extension of the standard argument\cite{algebra} leading to the quantization of
$k$.  Consider the $SU(2)$ subalgebras generated by (Weyl-Cartan basis)
\be
E^{1,0}_{\a} \quad ;\quad E^{-1,0}_{-\a} \quad ;\quad {2 \o {{\a}^2}}(\a .
H^{0,0} + k)
\ee
and
\be
E^{0,1}_{\a} \quad ;\quad E^{0,-1}_{-\a} \quad ;\quad {2 \o {{\a}^2}}(\a .
H^{0,0} + C_0 )
\ee
The unitary, highest weight representations of the algebra
(\ref{kma}-\ref{kmc}) will break into unitary, highest weight representations
of these subalgebras. Since ${2 \o {{\a}^2}}(\a .H^{0,0} + k)$ and  ${2 \o
{{\a}^2}}(\a . H^{0,0} + C_0)$ play the role of $2 T_3$, their eigenvalues must
be integers. But since for a given weight $\mu$, $2 \mu . \a /{{\a}^2}$ is
also an integer, one concludes that $k$ and $C_0$ must be integers (when the
squared length of the highest root is normalized to 2). Such an integral $C_0$
can be removed from the algebra, as we now show.

Consider the relabeling of the pair of indices $(m,r)$ defined by
\br
\( \begin{array}{c}
m'\\
r'
\end{array} \) = M
\( \begin{array}{c}
m\\
r
\end{array} \)
\label{transf}
\er
where $M = \( \begin{array}{cc}
a & b\\
c & d
\end{array} \) $; $a,b,c,d \in \IZ $ and $det M = 1$. The last condition is
needed in order to make the transformation one to one. We define then new
currents by
\be
\tilde{j}_a^{m,r} = \tilde{J}_a^{m',r'}
\ee
Using (\ref{kma}) and (\ref{transf}) the new currents satisfy the algebra
\br
\lb \tilde{j}^{m,r}_a  \, , \, \tilde{j}^{n,s}_b  \rb &=& i f_{ab}^c
\tilde{j}_c^{m+n,r+s}   +   k' g_{ab} r \d_{r+s,0}  \delta_{m+n,0}
+  {C'}_0   g_{ab} m \delta_{m+n,0} \d_{r+s,0}
\label{kmrot}
\er
with
\br
\( \begin{array}{cc}
{C'}_0 & k'
\end{array} \) =
\( \begin{array}{cc}
C_0 & k
\end{array} \) M
\er
and where we have used the fact that $\delta_{m'+n',0}
\d_{r'+s',0} =  \delta_{m+n,0} \d_{r+s,0}$ (if $detM \neq 0$).

We discuss separately the two possibilities:
\begin{enumerate}
\item $k$ is a divisor of $C_0$. So one can choose $a=1$, $b=0$, $c= -C_0/k$
and
$d=1$. Then $k'=k$ and ${C'}_0 = 0$. Analogously if $C_0$ is a
divisor of $k$ one can choose $a=k/{C_0}$, $b=1$, $c=-1$ and $d=0$. Then
one gets $k'=C_0$ and again ${C'}_0 = 0$.
\item If $C_0/k$ is rational, i.e., $C_0 = l p$ and $k = l q$ with $(p,q)$
relative primes, one chooses $a=q$ and $c=-p$ with $b$ and $d$ satisfying $dq +
b p = 1$, which always has a solution. One then gets $k'=l$ and
${C'}_0 = 0$.
\end{enumerate}
Consequently the last term on the r.h.s. of (\ref{kmrot}) can always be
transformed away reducing (\ref{kmrot}) to a very similar structure as given in
\cite{mai}.

Introducing the currents
\be
j^{r}_a (\bar x) \equiv {{1\over {2 \pi }}} \sum _{m=-\infty}^{\infty}
\tilde{j}^{m,r}_a e^{im{\bar x}}
\label{jotinha}
\ee
one can then write (\ref{kmrot}) as
\br
\lb j^{r}_a (\bar x)  \, , \, j^{s}_b (\bar y) \rb &=& \tilde{\delta}(\bar
x-\bar y) \( i f_{ab}^c
j_c^{r+s} (\bar x)   +   {k'\o {2 \pi}} g_{ab} r \d_{r+s,0}\)
\label{divkm}
\er
where $\tilde{\delta}(\bar x-\bar y) ={1\over 2\pi} \sum_{m=-\infty}^{\infty}
e^{im(\bar x-\bar y)} = \sum_{n=-\infty}^{\infty}\delta (\bar x-\bar y +2\pi
n)$. This indicates that the algebra is  decomposed into an infinity of
decoupled ordinary KM algebras and a decoupled $D-C$ system (except for their
zero modes which are still coupled to the ordinary KM's, having the role
mentioned above).

The highest weight representation of (\ref{divkm}) is constructed by applying
$j^r_a(\xb )$ on the highest weight state $\mid \lambda (\xb )>$ characterized
by
\br
j^r_a (\xb ) \mid \lambda (\yb ) > & =& 0 \, \, ; \, \, \, \, \, r>0
\nonumber \\
j^0_{\a } (\xb ) \mid \lambda (\yb ) > & =& 0
\er
where $\a $ is a positive root of the underlying finite Lie algebra, and
\be
j^0_{H_i} (\xb ) \mid \{ \lambda (\yb )\} > = \lambda_i (\xb ) \mid \{ \lambda
(\yb )\} >
\ee
where $j^0_{H_i} (\xb )$ correspond to the Cartan subalgebra generators. The
``distribution of weights'' $\lambda (\xb )$ is arbitrary, being limited by the
condition
\be
\int \lambda (\xb ) d \xb \equiv \lambda
\ee
where $\lambda $ is a weight of the ordinary Lie algebra.

For the decoupled $D-C$ system we have two options
\begin{enumerate}
\item A unitary representation, which however will not be highest weight, in
terms of the modes of a scalar field $\varphi$, $C(x)$ being identified with
$\partial_x \varphi$ and $D(x)$ with $\Pi _{\varphi}$ (the conjugate momentum).
\item A non unitary highest weight representation in terms of a $\beta -
\gamma$ system\cite{friedan}, $C(x)$ being identified with $\pa_x \beta$ and
$D(x)$ with $\gamma$.
\end{enumerate}

We remark that the delta function normalization in (\ref{divkm}) shows that
irreducible representations (irreps) of that algebra can not be constructed by
just taking infinite products of ordinary KM irreps. One can convince oneself
of that by considering the algebra (\ref{kmrot}) (with $C'_0=0$) with the
integers $m$ and $n$ varying from $-N$ to $N$ and the sum modulo $2N+1$. Making
a discret version of the transformation (\ref{jotinha}) one obtains
(\ref{divkm}) with a Kronecker delta and central term $k'\o {2N+1}$. Therefore
by taking the limit $N \rightarrow \infty $ one sees (\ref{kmrot}) is
equivalent to an infinity of ordinary KM algebras with vanishingly small
central extension.

We now show that the algebraic decomposition discussed above can be
implemented in the WZNW model having as current algebra the two loop
KM\cite{toda}. The arguments are not entirely rigorous but show the essential
features of the factorization.

The ordinary Lie algebra underlying the standard WZNW model is replaced by a
KM algebra with commutations relations
\br
\lbrack T^m_a \, , \, T^n_b \rbrack &=& if_{ab}^c T_c^{m+n} + \hc g_{ab}
m \delta_{m+n,0} \label{eq:affina}\\
\lbrack \hd \, , \, T^m_a \rbrack &=& m T^m_a  \label{eq:affinb} \\
\lbrack \hc \, , \, \hd \rbrack = 0 & \, \, \, &
\lbrack \hc \, , \, T^m_a \rbrack = 0 \label{eq:affind}
\er
and trace form
\br
\Tr ( T^m_a T^n_b ) =  g_{ab} \, \, \, ; \, \, \, \delta_{m+n,0}  \, \, \,
\Tr (\hc \hd ) = 1 \nonumber \\
\Tr (\hd \hd ) = \Tr (\hc \hc ) = \Tr (\hc T^m_a) = \Tr (\hd T^m_a) = 0
\label{killing}
\er
Defining formally a group element $\hg$ by
\be
\hg = exp \( i\sum_m T^m_a \a^a_{-m} + i\beta \hc + i\gamma \hd \)
\label{group}
\ee
The WZNW action is defined by
\be
S(\hg (x_{+},x_{-})) \equiv k \( {1\o {4 \pi}} \int_{\Sigma}
Tr(\hg^{-1}\pa_{+}\hg \hg^{-1} \pa_{-}\hg ) + {1\o {12 \pi}} \int_{B}
Tr(\hg^{-1} d\hg )^3 \)
\label{lag}
\ee
Since the global properties of the group defined by (\ref{group}) are not
known it is not clear at this stage if $k$ should be quantized and what are
the conditions on the zero modes of $\a$, $\beta$, and $\gamma$ which would
make $\hg$ single valued on the two dimensional manifold $\Sigma$. The
action (\ref{lag}) leads to the conservation of the left and right handed
currents $\pa_{-}\hg \hg^{-1}$ and $\hg^{-1} \pa_{+} \hg$. Their Poisson
brackets lead to the ``two-loop affine algebra'' for $\hat{J}(z) = \sum_{m,a,b}
g^{ab} J^m_a(z) T^{-m}_b + D(z) \hc + C(z) \hd $ where $z \in \Sigma$.

In order to facilitate some of the manipulations which will follow it is
convenient to present the algebra as  equal time commutators with spatial
variable $x$ being on a circle of radius 1 (i.e. $z=e^{ix}$) and the discrete
index $m$ with its conjugate variable $\xb \in (0,2\pi )$. This amounts to
defining
\be
J_a(x,\xb ) \equiv {1\o {2\pi}} \sum_{m=-\infty}^{\infty} J^m_a(x) e^{-im\xb }
\ee
The two-loop KM algebra (\ref{eq:kma})-(\ref{eq:kmd}) will be then
\br
\lb J_a (x,\xb ) \, , \, J_b (y,\yb ) \rb &=& i f_{ab}^c J_c(x,\xb ) \delta
(x - y) \delta (\xb -\yb )  + i k g_{ab} \pa_x  \d (x - y)
\delta (\xb -\yb ) \nonumber \\
& & + C (x) \d (x - y)  g_{ab} \pa_{\xb} \delta (\xb - \yb ) \phantom{......}
\label{contkma} \\
\lb D (x) \, , \, J_a(y,\yb ) \rb &=& i\pa_{\yb} J_a (y,\yb ) \d (x -y)
\label{contkmb} \\
\lb C (x) \, , \, D (y) \rb &=& i k \pa_x  \d (x - y)
\label{contkmc} \\
\lb C (x) \, , \, J_a (y,\yb ) \rb &=& 0 \label{contkmd}
\er
We return now to the action (\ref{lag}). A general group element $\hg $ can be
factorized, using the Baker-Campbell-Hausdorff theorem as follows
\br
\hg & \equiv & exp  \( i\int d\xb T_a (\xb ) \a^a (\xb ) + i\beta
\hc + i\gamma \hd \) \nonumber\\
& = & exp \( i\int d\xb T_a (\xb ) \rho^a (\xb ) \) exp  \( i\bar{\beta} \hc +
i\gamma \hd \) \equiv \hg_1 \hg_2
\label{baker}
\er
where $\rho^a (\xb )$ is the composite parameter, in the ordinary Lie group
sense (at each point $\xb$ separately) of the parameters $\sigma^a (\xb )$ and
$\sigma^a (\xb - \gamma )$; $\sigma^a$ being the solution of the differential
equation
\be
\pa_{\xb} \sigma (\xb ) = {M(\xb ) \o {\gamma (1-e^{M(\xb )})}} \a (\xb )
\ee
with the matrix $M(\xb )$ being defined as
\be
M_a^c (\xb ) = f_{ab}^c \a^b (\xb )
\ee
We do not reproduce the rather involved expression of $\bar{\beta}$ in terms of
$\a , \beta , \gamma$ which is not essential for our argument.
We use now the Polyakov-Wiegmann identity\cite{polyakov}
\be
S(\hg ) = S(\hg_1 ) + S(\hg_2 ) + {1\o {2\pi}}\int_{\Sigma} Tr\( \pa_{-} \hg_2
\hg_2^{-1} \hg_1^{-1} \pa_{+} \hg_1 \)
\label{polyakovka}
\ee
Notice that $\pa_{-} \hg_2 \hg_2^{-1}$ is a linear combination of the $\hd$ and
$\hc$ generators. Although $ \hg_1 $ is an exponential of a  linear combination
of $T_a (\xb )'s$ only, $\hg_1^{-1} \pa_{+} \hg_1 $ contains terms also in the
direction of $\hc$. So, according to (\ref{killing}), the last term on the
r.h.s. of (\ref{polyakovka}) apparently does not vanish. However we can make it
to vanish as follows. Write
\be
i\hg_1^{-1} \pa_{\mu} \hg_1 = \Gamma_{\mu} \hc + ...
\ee
Since $\Gamma_{\mu}$ is a vector in 2-D one can write it as
\be
\Gamma_{\mu} = \pa_{\mu} A + \epsilon_{\mu \nu} \pa^{\nu} B
\ee
Redefining $\hg$ in (\ref{baker}) as $\hg = \hg_1 \hg_2 =
\hg_1^{\prime } \hg_2^{\prime }$ with
\be
\hg_1^{\prime } = e^{i(A - B )\hc} \hg_1 \, \, ; \, \,
\hg_2^{\prime } = e^{-i(A - B )\hc} \hg_2
\ee
one gets
\br
i\hg_1^{\prime -1} \pa_{\mu} \hg_1^{\prime } & = & i\hg_1^{-1} \pa_{\mu} \hg_1
-\pa_{\mu} (A - B )\hc \nonumber\\
& = & (\pa_{\mu} B + \epsilon_{\mu \nu } \pa^{\nu} B ) \hc + ...
\er
It then follows that $\hg_1^{\prime -1} \pa_{+} \hg_1^{\prime } \sim
\hg_1^{\prime -1} \pa_{0} \hg_1^{\prime } +  \hg_1^{\prime -1} \pa_{1}
\hg_1^{\prime }$ does not contain any term in the direction of $\hc$.
Consequently the overlap term in (\ref{polyakovka}) will vanish.  Moreover due
to the properties of the trace form (\ref{killing}), $S(\hg_1^{\prime })$ does
not get any contribution of terms in the direction of $\hc$ and therefore it
could be equivalently generated by an algebra of $T^m_a$ with $\hc$ (and also
$\hd$) put to zero (so called loop algebra, and which is consistent with the
Jacobi identities being a direct product of decoupled ordinary Lie algebras at
each point $\xb$). Finally $S(\hg_2 )$ is simply
\be
S(\hg_2 ) = {k\o {4 \pi}} \int d^2x \pa_{\mu} \gamma \pa^{\mu} \bar{\beta}
\ee
i.e., a pair of scalar fieds coupled off-diagonally.

The current algebra that the two-loop decoupled actions (\ref{polyakovka})
would lead to is therefore
\br
\lb j_a (x,\xb ) \, , \, j_b (y,\yb ) \rb &=& \d (\xb - \yb ) \( i f_{ab}^c
j_c (x,\xb ) + i k g_{ab} \pa_x \d (x -y) \) \nonumber \\
\lb \tilde{C}(x) \, , \, \tilde{D}(y) \rb &=& i k \pa_x \d (x-y)
\label{decalg}
\er
all other commutators, including $\lb \tilde{D} \, , \, j_a \rb $ vanishing.

Now, since (\ref{lag}) and (\ref{polyakovka}) represent the same model, we
expect the algebras (\ref{contkma})-(\ref{contkmd}) and (\ref{decalg}) to be
equivalent. As we mentioned before the above argument is not rigorous since the
measure is not under control and  the zero modes may be shared between $\hg_1$
and $\hg_2$ leading to a failure of the factorization.

Let us now study the factorization from the point of view of
Chern-Simons-Witten (CSW) Lagrangeans.  As discussed in
ref.\cite{witten2,elitzur} a three dimensional CSW Lagrangean based on a Lie
algebra $\cal G$ has a gauge invariant Hilbert space related to the Hilbert
space of the two-dimensional WZNW model built on $\cal G$ and therefore to the
representation of the current algebra $\hat{\cal G}$.  Motivated by this
connection, in ref.\cite{bmz} a CSW Lagrangean based on the KM algebra
(\ref{eq:affina})-(\ref{eq:affind}) was proposed.  The gauge fields associated
with $T^{m}_{a}$, $\hc$ and $\hd$ respectively are $(A_{\mu})^{a}_{m}$,
$B_{\mu}$ and $C_{\mu}$.  It is convenient to replace the discrete index $m$ by
a continuous periodic variable $\bar {x} \in (0,2 \pi )$ through

\be
A^{a}_{\mu}(x,\bar x) =  \sum_{m \in \IZ}   (A_{\mu}(x))^{a}_{m} e^{ im{\bar
x}}
\ee

Then the CSW action proposed in \cite {bmz} can be rewritten as

\be
S = {1\over {2\pi}}\int _{M \otimes S^1}  d^{3}xd{\bar x} \epsilon ^{\mu \nu
\rho}({A^{a}_{\mu}\pa_{ \nu} A^{a}_{\rho} + {1\over
{3}}f_{abc}A^{a}_{\mu}A^{b}_{\nu}A^{c}_{\rho} + B_{\mu}(\pa _{\nu}C_{\rho} -
\pa _{\rho}C_{\nu}) + C_{\mu}{\pa A^{a}_{\nu}\over {\pa \bar x}} A^{a}_{\rho}})
\label {csaction}
\ee
where $M$ is a three dimensional manifold.  The gauge fields
$A^{a}_{\mu}(x,\bar x)$ should therefore be single valued on  $M \otimes
S^{1}$.  In the generating functional $Z$, which is given by
\be
Z = \int dAdBdC e^{ikS}
\label {genfunc}
\ee

\nit the integration over B leads to the delta function:
\be
\delta (\pa_{\nu}C_{\rho} - \pa_{\rho}C_{\nu})
\label {delta}
\ee

\nit This implies that
\be
C_{\rho } = \pa_{\rho}\alpha + {\theta_{\rho} \over {k}}
\label {soldelta}
\ee

\nit where $\theta_{\rho} $ are constants and $\alpha (x)$ is single
valued on M. Inside the generating functional (\ref {genfunc}) the
action $kS$ takes the form
\be
{k\over {2\pi}} \int d^{3}xd\bar x \epsilon^{\mu \nu \rho}(A^{a}_{\mu}\pa
_{\nu}A^{a}_{\rho} + {1\over {3}} f_{abc}A^{a}_{\mu}A^{b}_{\nu}A^{c}_{\rho} +
\p
   a
_{\mu}\alpha {\pa A^{a}_{\nu} \over {\pa \bar x}}A^{a}_{\rho} + {\theta _\mu
\over {k}}{ \pa A^{a}_{\nu} \over {\pa \bar x}}A^{a}_{\rho})
\label {newaction}
\ee

\nit Performing now the transformation
\be
(\bar A_{\mu}(x))^{a}_{m} = (A_{\mu}(x))^{a}_{m}e^{m\alpha (x)}
\label {abar}
\ee
\nit we obtain for the generating functional (\ref {genfunc})

\be
Z = \int d\bar A d\theta e^{ik\bar S} \int dBd \tilde C e^{ik \int d^{3} x
\epsilon ^{\mu \nu \rho}B_{\mu}(\pa _{\nu}\tilde C_{\rho} - \pa _{\rho}\tilde
C_{\nu}}) \label {seff}
\ee

\nit where $\tilde C_{\mu}$ stands for the $C_{\mu}$ field whithout the zero
mode
$\theta _{\mu}$ and where we have reinserted the integration over $B_{\mu}$

\be
k\bar S = {k \over {2\pi}} \int d^{3}x d\bar x \epsilon ^{\mu \nu \rho}(\bar
A^{a}_{\mu}\pa _{\nu}\bar A^{a}_{\rho} + {1\over {3}} f_{abc}\bar
A^{a}_{\mu}\bar
A^{b}_{\nu}\bar A^{c}_{\rho}) + {\theta _{\mu} \over {2 \pi}} \int d^{3}x d\bar
   x
\epsilon ^{\mu \nu \rho} {\pa \bar A^{a}_{\nu} \over {\pa \bar x}} \bar
A^{a}_\rho
\label{sbar}
\ee

\nit Notice that the Jacobian of the transformation (\ref {abar}) is one.  We
have
therefore shown that under the above transformations the generating functional
factorizes into two parts, one containing the $ B- \tilde C$ system and the
other
the $ \bar A-\theta$ system.

Let us now consider the transformation properties of $\bar S$ under the gauge
transformation

\be
\bar A_{\mu}(x,\bar x) \rightarrow g^{-1}(x,\bar x)\bar A_{\mu}g(x,\bar x) +
g^{-1}\pa _{\mu} g +{ \theta _{\mu} \over {k}} g^{-1} {\pa g(x,\bar x) \over
{\p
   a
\bar x}}
\label {gauge}
\ee

\nit Actually the action (\ref {sbar}) is automatically invariant under the
infinitesimal form of (\ref {gauge}).  Under the full finite transformation one
gets the additional terms

\br
{k \over {12 \pi}}f_{abc}\int_{M\otimes S^1}  d^{3}xd\bar x \epsilon ^{\mu \nu
\rho}(g^{-1}\pa _{\mu} g)^{a}(g^{-1}\pa _{\nu}g)^{b}(g^{-1}\pa _{\rho}g)^{c}
\nonumber \\
+ {\theta _{\mu} \over{4\pi}}f_{abc} \int_{M\otimes S^1} d^{3}xd\bar x \epsilon
^{\mu \nu
\rho}(g^{-1}{\pa g \over {\pa
\bar x}})^{a}(g^{-1}\pa _{\nu}g)^{b}(g^{-1}\pa _{\rho}g)^{c}
\label {wzterm}
\er
\nit If $g$ represents a nontrivial mapping from $M$ to the Lie group $G$  the
first integral does not vanish and in order that the path integral will be
invariant under $G$, we need to have $k$ quantized, i.e.$ k \in \IZ$.  In a
similar way choosing $M = S^{1}(x_{1}) \otimes M_{23}$ if $g(x_{\mu},\bar x)$
is a nontrivial mapping from the manifold $M_{23} \otimes S^{1}(\bar x)$ to $G$
we also need to quantize $\theta _{\mu }$, i.e. $\theta _{\mu } \in \IZ$.
Therefore $k$ and $\theta _{\mu}$ should be integers.  The situation becomes
particularly symmetric on a torus.  We shall now proceed in steps.  First we
perform a linear reparametrization  involving only coordinates $x^{\mu}$ by a
matrix $N$ with integral coefficients and determinant one.  Under such a
transformation the first two terms in (\ref {sbar}) are invariant, while the
third term induces a transformation on $\theta _{\mu}$ i.e.
\be
\theta _{\mu} ^{\prime} = (N^{-1})^{\nu} _{\mu} \theta _{\nu}
\label {theta}
\ee

\nit Using this transformation, $\theta _{\mu}$ can be oriented along one of
the directions, e.g. $\theta _{1} \not= 0,\theta _{2} = \theta _{3} = 0$.  We
next define a transformation of coordinates
\br
\( \begin{array}{c}
x'^1\\
\bar{x'}
\end{array} \) =
\( \begin{array}{cc}
a & b\\
c & d
\end{array} \)
\( \begin{array}{c}
x^1\\
\xb
\end{array} \) \, \, \, ; \, \, \, \, \, \, x'^2 = x^2, \, \, x'^3 = x^3
\label{matrix}
\er
with integral parameters $a, b, c, d$ and unit  determinant.  If $k$ and
$\theta_{1}$ have a maximal common divisor $l$, i.e. $k  = l \bar k$ and
$\theta_{1} = l \bar \theta_{1} $ with $\bar k$ and $\bar \theta_{1}$
being relatively primes, we define
\be
\tilde A_{1}(x_{\mu},\bar x) = \bar k \bar A_{1}(x_{\mu}^{\prime},\bar
x^{\prime
   })
\ee
\be
\tilde A_{2,3}(x_{\mu},\bar x) =  \bar A_{2,3}(x_{\mu}^{\prime},\bar
x^{\prime})
\ee
\nit with $a=\bar k$, $b=\bar \theta _{1}$ and $c,d$ are integral solutions of
$\bar{k} d  - \bar{\theta}_{1} c = 1$ which always exist.  Then the action
(\ref
{sbar}) can be rewritten as
\be
S={l \over {2 \pi}}\int d^{3}xd\bar x \epsilon ^{\mu \nu \rho}
(\tilde A^{a}_{\mu}\pa_{\nu}\tilde A^{a}_{\rho} +
 {1\over 3}f_{abc}\tilde A^{a}_{\mu}\tilde A^{b}_{\nu}\tilde A^{c}_{\rho})
\label{naction}
\ee
\nit i.e. the standard CSW model with $\bar x$ labelling decoupled gauge
fields.  A strange feature of the action (\ref {naction}) is the summation
over the coupling constant $l$ due to its implicit dependence on the $\theta
_{\mu}$ which are dynamical degrees of freedom.

The decoupled nature of (\ref{naction}) proves again our conjecture, namely,
the gauge invariant Hilbert space corresponding to (\ref {newaction}) is
related to the Hilbert space of decoupled WZNW models.  Therefore the Hilbert
space of the KM WZNW model in its original formulation (\ref {csaction}) is
isomorphic to the Hilbert space of decoupled ordinary WZNW models.

Finally let us comment on the possible applications of the above constructions
to the study of conformal models. As shown in ref.\cite{toda} the CAT model can
be obtained by Hamiltonian reduction from the WZNW model based on KM algebras.
It will be very interesting to investigate the consequences of the
factorization of the two-loop KM algebra to the CAT model.  This could perhaps
lead to a better understanding of CAT models and its relantionship to ordinary
Toda field theories.

{\bf Acknowledgements}
One of us (A.S.) would like to thank IFT-UNESP for the warm hospitality
extended to him.

\end{document}